\documentclass[aps,prc,groupedaddress,showpacs,twocolumn,floatfix]{revtex4}

\usepackage{graphicx}
\usepackage[dvips]{color}
\usepackage{colordvi}

\def\gtorder{\mathrel{\raise.3ex\hbox{$>$}\mkern-14mu
 \lower0.6ex\hbox{$\sim$}}}
\def\ltorder{\mathrel{\raise.3ex\hbox{$<$}\mkern-14mu
 \lower0.6ex\hbox{$\sim$}}}

\def\beq{\begin{equation}}
\def\eeq{\end{equation}}
\def\ba{\begin{eqnarray*}}
\def\ea{\end{eqnarray*}}
\newcommand{\rt}{\langle r \rangle_{(2)}}
\begin{document}

\title{Proton radii and two-photon exchange}

\author{Peter G.~Blunden}
\affiliation{Dept.~of Physics and Astronomy, University of Manitoba, 
Winnipeg, MB, Canada\ \ R3T~2N2}

\author{Ingo Sick}
\affiliation{Dept.~f\"{u}r Physik und Astronomie, Universit\"{a}t Basel,
CH4056 Basel, Switzerland}

\date{\today}

\begin{abstract}
We investigate the effect of two-photon exchange processes upon the
$rms$- and Zemach radii extracted from electron-proton scattering. We
find that the changes are small and do not help to explain the
discrepancy between experimental and calculated HFS in the hydrogen
atom.

\end{abstract}

\pacs{25.30.Bf, 21.20.Ft, 14.20.Dh}

\maketitle

\section{Introduction}
The recent progress in measurements of transition energies in the
Hydrogen atom has been remarkable. The 2p-1s transition energy and the
1s hyperfine structure interval (HFS) are now known to 14 and 12
significant digits, respectively \cite{Niering00,Eides01}. The
interpretation of these energies in terms of {\em e.g.} tests of QED now
depends entirely on the accuracy with which the proton finite-size
corrections are known.

These finite-size corrections can be determined from elastic
electron-proton scattering at low momentum transfer $q$. The proton
moment relevant for the 2p-1s energy (and the Lamb shift) is the charge
rms-radius $r_{rms}$ which has been extracted from the {\em world} data
on e-p scattering \cite{Sick03} to 0.895$\pm$0.018~fm. With this radius
as input, calculated and experimental 2p-1s transition energies agree
within the error bars. The proton moment relevant for the HFS is the
Zemach moment, derived from a convolution of charge- and magnetization
densities. This radius has also been extracted from the {\em world} e-p
data in Ref.~\cite{Friar04} which found $\rt=1.086\pm0.012$~fm. The
corresponding calculated HFS disagrees with experiment by 3.6(5)~ppm
\cite{Friar04} (see also \cite{Brodsky05a,Brodsky05b,Friar05}), where the uncertainty 
is dominated by the uncertainty in $\rt$.

This disagreement is partly explained by nuclear polarization effects in
the hydrogen atom, processes involving virtual excitation of the proton
to intermediary continuum states. Faustov {\em et al.} \cite{Faustov02}
have calculated this correction and find a contribution of 1.6~ppm in
the right direction using the experimental $g_1(q)$ and $g_2(q)$ spin
structure functions of the proton. The uncertainty of this correction is
hard to estimate as the nuclear polarization correction depends on $g_1,
g_2$ at very low $q$, where these functions are poorly known.

Before assigning the remaining discrepancy to this correction (or to not
yet calculated higher-order QED terms affecting the HFS) one should
note, however, that the moments extracted from electron scattering are
based on an interpretation of the data in one-photon exchange (plus
exchange of additional soft photons responsible for the Coulomb
distortion of the electron waves \cite{Sick98}). Electron scattering is
also subject to exchange of two hard photons, which {\em e.g.} have a
considerable effect upon the proton charge form factor $G_E$ as
determined from longitudinal/transverse separations at very large
momentum transfers \cite{Blunden03}.

In this Rapid Communication, we investigate the role of the two-photon
exchange in the determination of the proton moments, in order to find
out whether these corrections could be responsible for the discrepancy
with the HFS values.

\section{Calculation of two-photon exchange}
Details of the two-photon exchange correction to elastic electron-proton
scattering are described elsewhere \cite{Blunden03, Blunden05}. We
consider the contribution to the two-photon exchange box and crossed-box
amplitudes with an intermediate nucleon. At the low momentum transfers
of interest here (up to $q=4$~fm$^{-1}$) the contribution of an
intermediate $\Delta$ (or higher resonances) to the two-photon exchange
amplitude is negligible \cite{Kondratyuk05}.

Hadronic form factors consistent with the experimentally measured charge
and magnetic form factors $G_E(q^2)$ and $G_M(q^2)$ are introduced at
the photon-nucleon vertices. The two-photon exchange correction is
relatively insensitive to the particular choice of hadronic form factors
\cite{Blunden05}. For very low $q$, the two-photon exchange correction
behaves the same as that found for scattering from a point particle ({\em
e.g.} electron-muon scattering), and is therefore completely independent
of the hadronic form factors in this limit. This is a useful check on
our calculation.

\begin{figure}
\includegraphics[scale=0.45,clip]{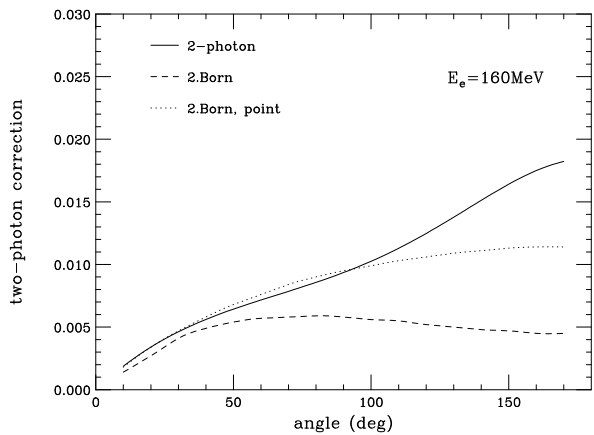}
\caption{Relative contribution of two-photon exchange to elastic e-p scattering at
E=160~MeV. The results in second Born-approximation account for the Coulomb distortion
(exchange of second soft photon) only.
\label{tdep}}
\end{figure}

In order to give an impression of the calculated results, we show in
Fig.~\ref{tdep} the two-photon correction for a typical electron energy.
The two-photon exchange contribution is compared to the contribution
involving only the piece from a second {\em soft} photon (Coulomb
distortion of the electron wave), calculated according to
Ref.~\cite{Sick98} in second Born approximation. The same contribution,
but for a point-nucleus, is given for comparison. Fig.~\ref{tdep} shows
that at forward angles the two-photon contribution is entirely dominated
by Coulomb distortion, while at backward angles the exchange of two hard
photons contributes appreciably.

\section{Analysis of world  e-p data}
It has been previously shown in Ref.~\cite{Sick03} that for an optimal
determination of the proton moments a parameterization in terms of a
Continued Fraction (CF) expansion should be used. For this
parameterization, the contribution of model-dependence has been
investigated in detail. As in Ref.~\cite{Sick03} we use the cross
sections up to a maximum momentum transfer of 4~fm$^{-1}$.

We use the {\em world} cross sections on e-p scattering (for references
see \cite{Sick03}). These data have been corrected for the contribution
of two-photon effects (omitting the second Born Coulomb corrections, which are
already included in the two-photon contribution), and then fitted with 
5-parameter CF expressions
for both the charge and the magnetic form factors $G_E(q^2)$ and $G_M(q^2)$.
The two-photon corrected longitudinal/transverse separation is thus done
implicitly during the fit. The charge-rms radius is obtained from the
slope at $q^2$=0 of $G_E(q^2)$, the Zemach moment is obtained via \ba
\langle r \rangle_{(2)} = -\frac{4}{\pi} \int_0^{\infty} \frac{d q}{q^2}
(G_E(q^2) G_M(q^2) - 1) ~~. \ea The statistical errors
have been determined using the error matrix, the systematic errors have
been obtained by changing the data sets individually by their systematic
errors, refitting the data and adding quadratically all the resulting
changes of the moments. The contribution of the model dependence is
accounted for as well. \\

\section{Results and conclusion}
The change of $\rt$ found when removing the contribution of two-photon
exchange (beyond Coulomb distortion) amounts to +0.0052~fm. This change
is small ($\sim$40\% of the error bar) and goes in the wrong direction
in terms of helping to explain the HFS discrepancy. The Zemach moment
with two-photon effects corrected for amounts to \ =1.091$\pm$0.012~fm.
The change in the charge-rms radius, +0.0015~fm, is also small. The
radius after two-photon correction amounts to 0.897$\pm$0.018~fm.

From these results we conclude that the discrepancy between calculated
and experimental HFS in the hydrogen atom cannot be attributed to
two-photon exchange contributions to (e,e) that could have falsified the
Zemach moment from e-p scattering. The origin of this discrepancy
presumably has to be sought in the uncertain nuclear polarization
correction to HFS, or potentially not yet calculated higher-order
contributions to the HFS.

\begin{acknowledgments}
This work has been supported by NSERC (Canada) and by the Schweizerische
Nationalfonds.
\end{acknowledgments}

\bibliographystyle{unsrt}
\bibliography{/usr/users/sick/sum2,/usr/users/sick/friar}


\end{document}